\documentclass[sigconf]{acmart}
\usepackage{physics}
\usepackage{algorithm}
\usepackage[noend]{algpseudocode}
\usepackage{graphicx}
\usepackage{subcaption}
\usepackage{booktabs} 

\begin{document}
\title{Investigating Order Effects in Multidimensional Relevance Judgment using Query Logs}

\author{Sagar Uprety}
\affiliation{%
  \institution{The Open University}
  \streetaddress{Walton Hall}
  \city{Milton Keynes}
  \country{United Kingdom}
  \postcode{MK7 6AA}
}
\email{sagar.uprety@open.ac.uk}

\author{Dawei Song}
\affiliation{%
  \institution{The Open University}
  \streetaddress{Walton Hall}
  \city{Milton Keynes}
  \country{United Kingdom}
} 

\additionalaffiliation{%
  \institution{Beijing Institute of Technology}
  \streetaddress{P.O. Box 1212}
  \city{Beijing}
  \country{China}
  \postcode{43017-6221}
}
\email{dawei.song@open.ac.uk}


\begin{abstract}
	There is a growing body of research which has investigated relevance judgment in IR being influenced by multiple factors or dimensions. At the same time, the Order Effects in sequential decision making have been quantitatively detected and studied in Mathematical Psychology. Combining the two phenomena, there have been some user studies carried out which investigate the Order Effects and thus incompatibility in different dimensions of relevance. In this work, we propose a methodology for carrying out such an investigation in large scale and real world data using query logs of a web search engine, and device a test to detect the presence of an irrational user behavior in relevance judgment of documents. We further validate this behavior through a Quantum Cognitive explanation of the Order and Context effects. 
\end{abstract}
%
%
\vspace{-6mm}


\copyrightyear{2018} 
\acmYear{2018} 
\setcopyright{acmcopyright}
\acmConference[ICTIR '18]{2018 ACM SIGIR International Conference on the Theory of Information Retrieval}{September 14--17, 2018}{Tianjin, China}
\acmBooktitle{2018 ACM SIGIR International Conference on the Theory of Information Retrieval (ICTIR '18), September 14--17, 2018, Tianjin, China}
\acmPrice{15.00}
\acmDOI{10.1145/3234944.3234972}
\acmISBN{978-1-4503-5656-5/18/09}
\maketitle
\vspace{-0.6mm}
\section{Introduction}
The concept of relevance in Information Retrieval (IR) has been shown to consist of different dimensions~\cite{ASI:Barry-relevance, ASI:Tombros-relevance, ASI:Xu-relevance,  MURM-psychometrics, ASI:Xu-MURM, ASI:Jingfei,10.1007/978-3-642-00958-7_25}. In a recent work, ~\cite{ASI:Jingfei} extended the multidimensional user relevance model (MURM) proposed by \cite{ASI:Xu-MURM,MURM-psychometrics}. The extended MURM includes seven dimensions of relevance, namely "Habit", "Interest", "Novelty", "Reliability", "Scope", "Topicality" and "Understandability", each of which is quantified for a document-query pair by a set of features extracted from the document or the document-query pair~\cite{ASI:Jingfei}. On the other hand, ~\cite{Bruza10.3389/fpsyg.2014.00612} investigated possible interactions among relevance dimensions in terms of their incompatibility, thus giving rise to the Order Effects in relevance judgment. Order Effects in decision making based on incompatible perspectives are being actively investigated in the recent field of Quantum Cognition~\cite{10.1007/978-3-642-00958-7_25,Aerts2009,Pothos2009,Busemeyer2011,Trueblood2011,Busemeyer:2012:QMC:2385442}. It uses the mathematical formalism of Quantum Theory to model, explain and predict irrational human behavior.

In this paper, we propose to use large scale query logs of real world search engines to investigate such behavior in multi-dimensional relevance judgment. A simple test for detecting irrational behavior is presented. The counter-intuitive results obtained are explained using the Quantum Cognitive model of Order Effect between the different relevance dimensions. The incompatible relevance dimensions are represented as different basis of a Hilbert space, based on the Hilbert space construction method devised in \cite{sigir18}.

\vspace{-3mm}
\section{Order effects in relevance judgment}
A fundamental assumption of the classical probability theory is that conjunction of two events is commutative. That is, $P(A \cap B ) = P(B \cap A)$. Thus one can always form a joint probability distribution to assign probabilities to conjunction of events. However in some cases, human decision making for multiple events has been found to be non-commutative~\cite{Hogarth1992,Trueblood2011}. For example, consider a case where a user is asked to judge an online forum post based on two decision perspectives: (a) Is the post relevant to the topic of discussion (denoted as $T$) and (b) Is the sentiment of the post positive (denoted as $+$). Essentially one has to calculate $P(T, +)$. Assume that the user's cognitive state with respect to the post is uncertain, with the user being 90\% certain that the post is topically relevant and 40\% certain that the post has a positive sentiment. In the formalism of Quantum Theory, we represent the user's cognitive state as a vector (denoted as $\ket{S}$) in a two dimensional Hilbert space. Positive and negative sentiments are orthogonal vectors which span the Hilbert space. Also, Topical and Not-topical are another set of orthogonal vectors in the same Hilbert space~(Figure 1.a.). These two basis represent two different ways of evaluating the forum post - one from the topical perspective, and another from the sentiment perspective. Thus:
\vspace*{-3mm}
\begin{align}
\ket{S} &= 0.9487\ket{T} +0.3162\ket{\widetilde{T}}\\ \nonumber
 		  &= 0.6325\ket{+} +0.7746\ket{-}
\end{align}
where $\ket{T}$ and $\ket{\widetilde{T}}$ represent the vectors for the post being Topical and not topical respectively. The notation $\bra{T}\ket{S}$ is the inner product of the vectors $T$ and $S$. It denotes the probability amplitude, which is in general, a complex number. For real valued vectors, inner product is same as the dot product. According to the Born rule for calculating Quantum probabilities, the probability that the post is topically relevant for the given user state, is the square of the projection of the user state $\ket{S}$ on the vector for Topicality $\ket{T}$, which gives us $|\bra{T}\ket{S}|^2 = 0.9487^2 = 0.90$.

Using the derivation in Appendix A, we also get:
\vspace{-2mm}
\begin{align}
\ket{+} &= 0.8449\ket{T} - 0.5349\ket{\widetilde{T}}
\end{align}

\begin{figure*}[h!]
    \begin{subfigure}[t]{0.288\textwidth}
        \includegraphics[width=\textwidth]{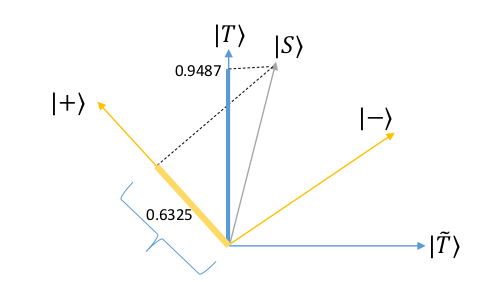}
        \caption{Figure 1.a}
    \end{subfigure}
    \begin{subfigure}[t]{0.288\textwidth}
        \includegraphics[width=\textwidth]{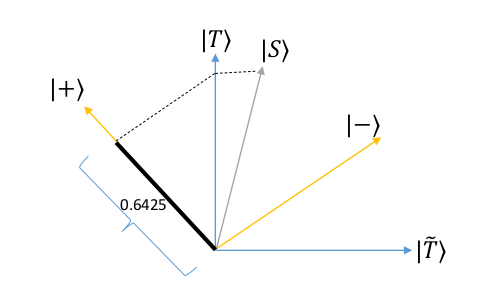}
        \caption{Figure 1.b}
    \end{subfigure}
        \begin{subfigure}[t]{0.288\textwidth}
        \includegraphics[width=\textwidth]{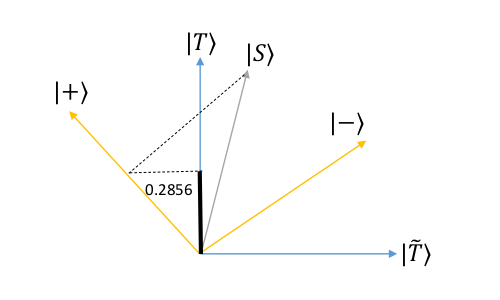}
        \caption{Figure 1.c}
    \end{subfigure}
\end{figure*}

Now we calculate the probability of the post being topically relevant and of positive sentiment in two ways - (a) The user evaluates the post first considering its topicality and then the sentiment, (b) The user first considers the sentiment of the post and then the topicality. For case (a), we take the user's cognitive state and first project it onto the vector for Topicality($\ket{T}$), and then project the resulting state onto the vector for positive sentiment. Thus we take the path $\ket{S} -> \ket{T} -> \ket{+}$~(Figure 1.b). We get, 
\begin{align}
P(T, +) = |\bra{T}\ket{S}|^2.|\bra{+}\ket{T}|^2 \nonumber
		&= 0.9487^2 * 0.8449^2 = 0.6425
\end{align}

For the path  $\ket{S} -> \ket{+} -> \ket{T}$~(Figure 1.c), we get,
\begin{align}
P(+, T) = |\bra{+}\ket{S}|^2.|\bra{T}\ket{+}|^2 
		&= 0.6325^2 * 0.8449^2 = 0.2856
\end{align}

Thus we see that user's judgment regarding the post being positive and topical exhibits an Order Effect. Different order of consideration leads to different judgment. Geometrically, it arises from defining the two decisions of topic and sentiment as different basis. In the Quantum Framework, if two events do not share a common basis, they are incompatible and thus they have to be evaluated sequentially. A single joint representation does not exist. The Quantum Framework has been successful in modeling, and predicting such order effects in real world data~\cite{Trueblood2011} and we make use of it to detect Order Effects in relevance judgment in Information Retrieval.

\vspace{-3mm}

\section{Test of Order Effects in Relevance Judgment With Query Logs}
In \cite{Bruza10.3389/fpsyg.2014.00612}, participants were asked to judge documents based on different dimensions of relevance. For one document, there were two sets of relevance judgments, for two different orders of relevance dimensions. Hence, relevance probabilities could be calculated based on different orders of relevance dimensions, which could be tested using the q-test\cite{Wang2013}. Such user study based approach, on the one hand, directly captures users' responses in a controlled environment, while on the other hand, is limited in terms of scalability and reflection of natural and real-world IR settings. In this paper, we propose to investigate Order Effects in multi-dimensional relevance judgment with query logs from real-world search engines. Specifically, a Bing query log dataset is used in our study.
\vspace{-3mm}
\subsection{Dimensional Profile}
We model documents in terms of the seven relevance dimensions mentioned earlier, and propose the concept of ``Dimensional Profile'' of a document that represents the relevance criteria used for judging the document with respect to a specific query. Consider each document to be a seven dimensional vector, where each value in the vector corresponds to the importance of the corresponding relevance dimension to the document. This vector forms the Dimensional Profile of the document. 

The similarity between two documents based on their Dimensional Profiles can be measured based on the relative difference between the values of each dimension for the two documents. For example, if the Dimensional Profile of document $d_1$ is given by the vector $[\alpha_{11}, ..., \alpha_{17}]$ and that of document $d_2$ given by $[\alpha_{21}, ..., \alpha_{27}]$, the relative difference vector is\\
$[\frac{|\alpha_{21} - \alpha_{11}|}{max(\alpha_{21}, \alpha_{11})}, ...., \frac{|\alpha_{27} - \alpha_{17}|}{max(\alpha_{27}, \alpha_{17})}]$. 
We then specify a matching criteria where each value in the difference vector should be within that criteria. For example, matching criteria of value $0.05$ means that the differences in corresponding scores for the dimensions between the two documents are all within $5\%$. A matching criteria of $0$ means both the documents have exact same scores for all the seven dimensions. In Section 4 we show how these scores are calculated from the query log data.

\vspace{-3mm}
\subsection{Irrational User Behavior}
With query log data, we have only one relevance decision per document. Hence it is not possible to test judgments based on different orders of relevance dimensions. As such, using the q-test to detect Order Effects and thus incompatibility is not possible. We test for these effects in a different way, as detailed below.

First, we identify the subset of queries where the first two retrieved documents have a similar Dimensional Profile. We name this subset as $\textbf{similarFirstTwo (SFT)}$. Next, from this subset we find out those queries where second document is SAT clicked\footnote{Those documents which are clicked and browsed for over 30 seconds}. This subset is named as $\textbf{similarFirstTwoSecondClicked (SFTSC)}$. Now, if the second document is clicked and its dimensional profile is similar to the first, rationally one would expect the user to have SAT-clicked the first document as well. Not doing so would suggest an irrational judgment. Hence, among the queries of the set $\textbf{SFTSC}$, we find if there are queries where the first retrieved document is not SAT-clicked. We indeed find that there are such queries and we label this subset as $\textbf{irrationalBehaviorQueries (IRQ)}$. 

Out of the total $152941$ queries in the query log dataset used in this study, we found 170 queries where the top two documents have the same Dimensional Profile (Matching criteria of 0). Of these $170$, in $25$ queries we had only second document SAT clicked and not the first one. Although $\textbf{SFT}$ does not form a significant fraction of the total queries analyzed, they represent the set of those queries which can potentially exhibit user's irrational behavior. The subset $\textbf{IRQ}$ forms $14.71\%$ of the subset $\textbf{SFT}$, which is a significant fraction. Table 1 summarizes the results with different criteria of Dimensional Matching. 

\begin{table}[h!]
\begin{center}
\resizebox{\columnwidth}{!}{%
 \begin{tabular}{||c|c|c|c|c||} 
 \hline
Matching Criteria & SFT & SFTSC & IRQ & IRQ percent(of SFT) \\ [0.1ex] 
 \hline\hline
 10\% & 309 & 44 & 40 & 12.94 \\ 
\hline
5\% & 238 & 30 & 27 & 11.34 \\
\hline
\textbf{0}\% & \textbf{170} & \textbf{27} & \textbf{25} & \textbf{14.71} \\
\hline
 \end{tabular}
 }

\end{center}
\caption{Analysis of Bing query log dataset}
\end{table}
\vspace*{-10.5mm}
\section{Quantifying Dimensions of Relevance}
We represent each document as a two-dimensional real valued Hilbert space. The two basis vectors correspond to relevance and non-relevance of a dimension. For the seven dimensions, we have seven different basis in the Hilbert space. The user's cognitive state for this document is a vector in the Hilbert space, a superposition of the basis vectors. Using the Dirac notation, we get the user state for a document d as:
\vspace{-2mm}
\begin{align}
\ket{d} &= \alpha_{1}\ket{Habit} + \beta_{1}\ket{\widetilde{Habit}} \nonumber \\
	      &= \alpha_{2}\ket{Interest} + \beta_{2}\ket{\widetilde{Interest}}
\end{align}
and so on, in all seven basis. The coefficients $|\alpha_x|^2$ is the weight (i.e., probability of relevance) the user assigns to document $d$ in terms of the dimension $x$ , and $|\alpha_x|^2 + |\beta_x|^2 = 1$. The different basis correspond to the different perspectives of relevance judgment for the document. Based on which dimension is considered, the same document will have different relevance probabilities. Thus the document exists in multiple states (e.g. highly relevant, not relevant, moderately relevant, etc.) simultaneously and we get a particular judgment depending upon which criteria (relevance dimension) the user used to judge (measure) it. This is analogous to the measurement of electron spin which is either up or down in direction, but depends upon which axis it is measured in. Electrons with spin up along the Z-axis may have both up and down components along the X-axis. So a document may look relevant based on the Topicality dimension, but may not be so along, say, the Reliability dimension.

To calculate the coefficients of superposition in a basis, we use the same technique as \cite{sigir18}. Following the methodology in \cite{ASI:Jingfei}, we define a set of features for each of the seven relevance dimensions. For each query-document pair, the set of features for each dimension are extracted and integrated into the LambdaMART~\cite{Burges2010FromRT} Learning to Rank (LTR) algorithm to generate seven relevance scores (one for each dimension) for the query-document pair. Due to lack of space, we refer the readers to \cite{ASI:Jingfei} for more details on the features defined for each dimension and also how they are used in the LTR algorithm. We thus get seven different ranked lists for a query, corresponding to each relevance dimension. Then the scores assigned to a document for each dimension are normalized using the min-max normalization technique, across all the documents for the query. The normalized score for each dimension forms the coefficient of superposition of the relevance vector for the respective dimension. For example, for a query $q$, let $d_1, d_2, ..., d_n$ be the ranking order corresponding to the "Scope" dimension, based on the relevance scores of $\lambda_1, \lambda_2, ..., \lambda_n$ respectively. We construct the vector for document $d_1$ in the 'Scope' basis as:
\begin{equation}
\ket{d_1} = \alpha_{11}\ket{Scope} +\beta_{11}\ket{\widetilde{Scope}}
\end{equation} 
where $\alpha_{11} = \sqrt{\frac{\lambda_1 - min(\lambda)}{max(\lambda) - min(\lambda)}}$, where $max(\lambda)$ is the maximum value among $\lambda_1, \lambda_2, ..., \lambda_n$. Square root is taken to enable calculation of probabilities according to the Born rule of Quantum Physics. We can thus represent this document in all the seven basis and therefore all the documents in their respective Hilbert spaces.

The Dimensional Profile vector of a document is made up of the normalized scores for the seven dimensions of the document. That is, for the above document $d_1$, it will be $[\alpha_{11}^2, \alpha_{12}^2 ..., \alpha_{17}^2]$.

\vspace{-1mm}
\section{Quantum Cognitive Explanation of Irrational Relevance Judgments}

\begin{figure*}[t!]
    \begin{subfigure}[t]{0.28\textwidth}
        \includegraphics[width=\textwidth]{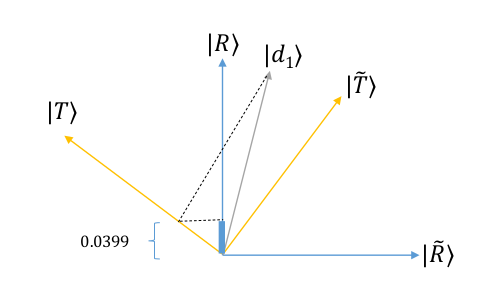}
        \caption{Figure 2.a}
    \end{subfigure}
    \begin{subfigure}[t]{0.28\textwidth}
        \includegraphics[width=\textwidth]{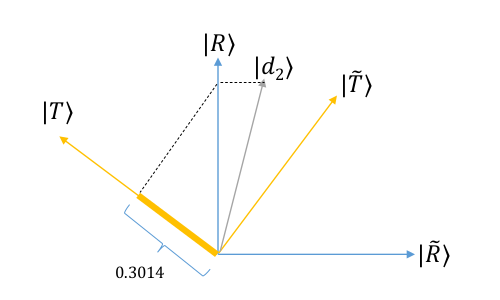}
        \caption{Figure 2.b}
    \end{subfigure}
\end{figure*}
Having obtained results as shown in Table 1, we take one query from set $IRQ$ and analyze the Hilbert spaces of the first two documents. Table 2 shows the user state vector of one such query with respect to the relevance vectors for the seven dimensions of Habit, Interest, Novelty, Reliability, Scope, Topicality, Understandability (\textbf{HINRSTU}) respectively. 

Here we see that for both documents, the scores of all dimensions are exactly the same and the second document is SAT-clicked. We hypothesize that the dimension with the highest score is the preferred dimension for the query, which is $Reliability$ in this case. Also, we can see that there are a few dimensions with very low scores. Let us construct a Hilbert space for Document 1 representing the basis for $Reliability$ and $Topicality$. So we have:
\vspace{-2mm}
\begin{align}
\ket{d_1} &= 0.9715\ket{Reliability} +0.2370\ket{\widetilde{Reliability}}\\
 		  &= 0.3535\ket{Topicality} +0.9354\ket{\widetilde{Topicality}}
\end{align}
where, $0.9715 = \sqrt[]{0.9438}$ and so on. We take the $Reliability$ basis as the standard basis. Representing $Topicality$ basis in the standard $Reliability$ basis, we get (Appendix A):
\begin{equation}
\ket{Topicality} = 0.5651\ket{Reliability} + 0.8250\ket{\widetilde{Reliability}}
\end{equation}

Suppose that while judging Document 1, the user has the order $Topicality$ -> $Reliability$ in mind. Then the final probability of relevance is the projection from $d_1$->$T$->$R$ as shown in Figure 2.a. This is calculated as $|\bra{T}\ket{d_1}|^2|\bra{R}\ket{T}|^2 = 0.3535^2*0.5651^2 = 0.0399$. If the user reverses the order of relevance dimensions considered while judging document $d_2$, we get $d_2$->$R$->$T$ = $|\bra{R}\ket{d_2}|^2|\bra{T}\ket{R}|^2 = 0.9715^2*0.5651^2 = 0.3014$, which is $7.5$ times larger (Figure 2.b). Now since the Hilbert space for both Document 1 and Document 2 are same, we get these results if the user follows the order $d_1$->$T$->$R$ for Document 1 and $d_2$->$R$->$T$ for Document 2. Hence Document 2 ends up being SAT-clicked.

The important question to ask here is what causes the user to use two different orders for the two documents. To explain this, we suspect an Attraction effect as a type of Context effect~\cite{ASI:Order}. Since the user considers $Reliability$ as the last dimension when judging Document 1, there is a memory bias leading to an Attraction effect towards $Reliability$, and thus the user context changes to consider it first while judging the second document.

\begin{table}[h!]
\begin{center}
\resizebox{\columnwidth}{!}{%
 \begin{tabular}{||c|c|c|c|c|c|c|c||} 
 \hline
Document rank & H & I & N & R & S & T & U \\ [0.1ex] 
 \hline\hline
1 & 0.3040 & 0.1251 & 0.0000 & 0.9438 & 0.1250 & 0.1250 & 0.5619 \\
\hline
2 & 0.3040 & 0.1251 & 0.0000 & 0.9438 & 0.1250 & 0.1250 & 0.5619 \\
\hline
 \end{tabular}
 }
\end{center}
\caption{Random Query Analysis}
\end{table}
\vspace*{-12mm}
\section{Conclusion and Future Work}
In this work we have found evidence of Order Effects in relevance judgment based on multiple dimensions. It is empirically shown how the incompatible representation of relevance dimensions leads to these Order Effects. The number of queries showing such effects might be very small, but note that we have not tested for all the different possibilities of Order Effects. We take only those queries where the first and second documents had similar dimensional profiles. There might be Order Effects exhibited in relevance judgment in other documents down the ranking list. In the future we intend to investigate all possible situations where such effects might occur. Also, we would like to find out if detecting Order Effects can help in improving document ranking. At present, the Order Effect takes place at the cognitive level of the user. Only if we know what dimensions the user prefers for a given query or task, can we predict the Order Effects and rank the documents accordingly. The preference for relevance dimensions can be either dependent on the user's personality or the type of query/task or both. Therefore we can detect the dimensional preference by either building a cognitive profile based on user's previous history, or classify queries/tasks based on the what relevance dimensions users consider as important for those queries/tasks. 
\vspace{-3mm}
\appendix
\section{Appendix}
Consider a state vector in two different basis of a two dimensional Hilbert space, $\ket{\psi} = a\ket{A}+b\ket{B} = c\ket{C}+d\ket{D}$
We want to represent the vectors of one basis in terms of the other. To do that, consider the vector orthogonal to $\ket{\psi}$, which is $\ket{\widetilde{\psi}} = b\ket{A}-a\ket{B} = d\ket{C}-c\ket{D}$. Using the above representations, we get 
\vspace{-2mm}
\begin{equation}\label{c,d}
\ket {C} = c\ket{\psi} + d\ket{\widetilde{\psi}} and 
\ket {D} = d\ket{\psi} - c\ket{\widetilde{\psi}}
\end{equation}
\vspace{-2mm}
Substituting $\ket{\psi} = a\ket{A}+b\ket{B}$ and $\ket{\widetilde{\psi}} = b\ket{A}-a\ket{B}$ in \ref{c,d}, we get:
\begin{align}
\ket {C} &= (ac+bd)\ket{A} + (bc-ad)\ket{B} \nonumber \\
\ket {D} &= (ad-bc)\ket{A} + (ac+bd)\ket{B}
\end{align}
\vspace*{-8mm}
\begin{acks}
This work is funded by the European Union's Horizon 2020 research and innovation programme under the Marie Sklodowska-Curie grant agreement No 721321.
\end{acks}
\bibliographystyle{ACM-Reference-Format}
\vspace*{-4mm}
\bibliography{bibliography}

\end{document}